\documentclass[12pt]{book}

\usepackage[dvips]{graphicx,color}
\usepackage{times,makeidx,view}
\makeauthorindex
\makeindex

\begin{document}

\BookTitle{\itshape}
\CopyRight{\copyright 2002 by Universal Academy Press, Inc.}
\pagenumbering{arabic}

\chapter{The VERITAS Project}

\author{R.A.~Ong$^{1}$, 
I.H.~Bond$^{2}$,
P.J.~Boyle$^{3}$,
S.M.~Bradbury$^{2}$, 
J.H.~Buckley$^{4}$,
I.~de~la~Calle~Perez$^{2}$, 
D.A.~Carter-Lewis$^{5}$,
O.~Celik$^{1}$,
S.~Criswell$^{6}$,
W.~Cui$^{7}$, 
M.~Daniel$^{5}$,
A.~Falcone$^{7}$,
D.J.~Fegan$^{8}$, 
S.J.~Fegan$^{6}$,
J.P.~Finley$^{7}$,
L.F.~Fortson$^{3}$,
J.A.~Gaidos$^{7}$, 
S.~Gammell$^{8}$,
K.~Gibbs$^{6}$, 
D.~Hanna$^{9}$,
J.~Hall$^{10}$,
A.M.~Hillas$^{2}$, 
J.~Holder$^{2}$,
D.~Horan$^{6}$, 
M.~Jordan$^{4}$,
M.~Kertzman$^{11}$,
D.~Kieda$^{10}$, 
J.~Kildea$^{9}$,
J.~Knapp$^{2}$,
K.~Kosack$^{4}$, 
H.~Krawczynski$^{4}$,
F.~Krennrich$^{5}$,
S.~LeBohec$^{5}$, 
E.~Linton$^{3}$,
J.~Lloyd-Evans$^{2}$,
P.~Moriarty$^{12}$,  
D.~M\"uller$^{3}$,
T.~Nagai$^{10}$, 
S.~Nolan$^{7}$,
R.~Pallassini$^{2}$,
F. Pizlo$^{7}$,
B.~Power-Mooney$^{8}$,
J.~Quinn$^{8}$, 
K.~Ragan$^{9}$,
P.~Rebillot$^{4}$,
J.~Reynolds$^{13}$,
H.J.~Rose$^{2}$, 
M.~Schroedter$^{6}$,
G.H.~Sembroski$^{7}$, 
S.P.~Swordy$^{3}$, 
V.V.~Vassiliev$^{10}$, 
S.P.~Wakely$^{3}$,
G.~Walker$^{10}$, 
T.C.~Weekes$^{6}$,
J.~Zweerink$^{1}$}

\author{$^{1}$Dept. of Physics \& Astron., Univ. of California,
Los Angeles, CA 90095, USA\\
$^{2}$Dept. of Physics, University of Leeds, Leeds, LS2 9JT, Yorkshire, UK\\
$^{3}$Enrico Fermi Inst., University of Chicago, Chicago, IL
60637, USA\\
$^{4}$Dept. of Physics, Washington University, St.~Louis, MO
63130, USA\\
$^{5}$Physics \& Astronomy Dept.,
Iowa State University, Ames, IA 50011, USA\\
$^{6}$F.~Lawrence Whipple Obs., Harvard-Smithsonian CfA,
Amado, AZ 85645, USA\\
$^{7}$Dept. of Physics, Purdue University, West Lafayette, IN
47907, USA\\ 
$^{8}$Dept. of Physics, National University of Ireland,
Dublin 4, Ireland\\
$^{9}$Dept. of Physics, McGill University, 
Montreal, Quebec H3A 2T8, Canada\\
$^{10}$High Energy Astrophysics Inst., Univ. of Utah, Salt
Lake City, UT 84112, USA\\
$^{11}$Physics Dept., De Pauw University, Greencastle, 
IN 46135, USA\\
$^{12}$School of Science, Galway-Mayo Institute of Technology,
Galway, Ireland\\
$^{13}$Dept. of Applied Physics, Cork Institute of Technology, Cork, 
Ireland\\
}

\AuthorContents{R.A. Ong, VERITAS Collaboration} 

\AuthorIndex{Ong}{R.A.}

\section*{Abstract}

VERITAS is a new, major ground-based $\gamma$-ray observatory
designed to significantly advance our understanding of
extreme astrophysical processes in the universe.
The observatory comprises seven large-aperture (12$\,$m diameter)
Cherenkov telescopes, each equipped with an imaging camera.
The first phase of VERITAS (consisting of four telescopes)
is currently under construction.
Here we outline the key features of VERITAS and provide an
update on its status [1].

\section{Introduction}

Ground-based $\gamma$-ray astronomy
came of age in the last decade.
The discovery of 
numerous sources by telescopes using
the imaging atmospheric Cherenkov technique proved that
very high energy (VHE) astronomy can be done from the ground.
Roughly one dozen VHE sources have now been detected 
with varying levels of significance [2].
The established sources include 
pulsar-powered nebulae,
active galaxies of the BL Lac type, and
shell type supernova remnants.

These exciting discoveries strongly motivate the construction
of new instruments to substantially increase the VHE source
catalog and to enable much more detailed studies of
individual sources.
VERITAS combines the successful features
of the Whipple Observatory (large-aperture reflector,
imaging camera) with those of HEGRA
(array of reflectors).  Relative to current operating imaging telescopes,
VERITAS will have substantially {\it better flux sensitivity}
(factor of five to ten improvement, depending on energy),
{\it reduced energy threshold} (peak energy near 100 GeV),
{\it improved energy resolution} (resolution of 10-15\% over
a broad energy range), and {\it improved angular resolution}
(4.3 arc-min at 1 TeV).

\section{VERITAS Design}

Serious consideration of VERITAS started in 1996,
and its design was developed and finalized over the next several
years.
The design uses modular construction and
proven technology to a large degree, but it also
encompasses new technical innovations where appropriate.
VERITAS will be an array of seven telescopes, each employing a $12\,$m
diameter optical reflector.
Each telescope has a camera of 499 photomultiplier tube elements, covering
a field of view of 3.5$^\circ$ diameter.
The photomultiplier tubes are
read out through high-bandwidth 
electronics by a Flash ADC (FADC) system sampling at 500$\,$MSps.
The FADC sampling, a flexible trigger system, and extensive
electronic and optical calibration systems are among the
important new capabilities of VERITAS.
Details on the design and expected performance of VERITAS have been
published earlier [3,4].  
This report provides an update on the key developments that have happened 
recently.

\begin{table}
\vspace*{-0.3in}
\begin{center}
\caption{VERITAS-4 Performance}
\begin{tabular}{rcl}
\hline\hline
Characteristic & \qquad\qquad E\qquad\qquad\qquad & Value 
\\ \hline
Peak Energy $^a$ & & 110 GeV \\
Flux sensitivity & 100\,GeV & 3.4$\times$10$^{-11}$cm$^{-2}$s$^{-1}$ \\
 & 1\,TeV & 6.5$\times$10$^{-13}$cm$^{-2}$s$^{-1}$ \\
 & 10\,TeV & 2.1$\times$10$^{-13}$cm$^{-2}$s$^{-1}$ \\
Angular resolution & 100\,GeV & 7.5 arc min \\
 & 1\,TeV & 4.3 arc min \\
 & 10\,TeV & 1.6 arc min \\
Collection area & 100\,GeV & 3.3$\times$10$^8$cm$^2$ \\
 & 1\,TeV & 2.2$\times$10$^9$cm$^2$ \\
Crab Nebula & $>$100\,GeV & 40/minute \\
$\gamma$-ray rates & $>$300\,GeV & 15/minute \\
 & $>$1\,TeV & 4/minute \\
Energy resolution$^b$  & & $<$15\% \\ \hline
\multicolumn{3}{p{5.5in}}{\small$^a$Maximal differential rate for
Crab Nebula-like spectrum.
The trigger requirement is three adjacent pixels per telescope, each with
more than 5.6 p.e. and three out of four telescopes.
}\\
\multicolumn{3}{p{5.5in}}{\small$^b$RMS $\Delta E/E$.} \\
\end{tabular}
\end{center}
\end{table}

VERITAS will be sited on Mt. Hopkins in southern Arizona, USA, near
the Basecamp of the Whipple Observatory  
(1350$\,$m altitude).
The site will be well protected from light pollution
(within 20\% of the darkest sites available in
the U.S.).
The construction will be carried out in two phases.
During the first phase (VERITAS-4), four telescopes will be constructed
and deployed as shown in Figure~1.
In the second phase, three additional telescopes will be 
added to complete the hexagonal array.

\section{VERITAS-4 Performance}

We have carried out detailed simulation studies to estimate the expected
performance of the four telescope array, VERITAS-4 -- some
key results are shown in Table~1.
In quoting sensitivity, we make the conservative requirement
of at least a five standard deviation $\gamma$-ray excess 
in each energy bin, of width one-quarter decade.

\section{Technical Progress}

A great deal of technical progress has been made on VERITAS, and
we are well on our way towards the construction
of a prototype telescope encompassing all important design
elements.  
Here we describe the salient features of the VERITAS design and
recent progress made on the construction of the prototype
that will become operational in mid-2003.

\subsection{Telescope, Optical Support Structure, and Mirrors}

The VERITAS telescopes consist of a tubular-steel, space-frame
optical support structure (OSS) mounted on a commercial positioner.
In the OSS design, as shown in Figure~1,
quadrapod arms penetrate the mirror surface to hold up the camera.
We expect excellent optical performance from the OSS --
the blur will be well less than $0.01^\circ$ over the
usable elevation range of the telescope, and
the de-centering will be less than $0.002^\circ$.
The positioner for the prototype telescope will be delivered in
April 2003.

We use the Davies-Cotton optical design
in which the telescope reflecting surface is spherical and
the mirror facets are identical in shape.
Each $12\,$m diameter (f/1.0) mirror 
is made from 350 hexagonal facets and has a total mirror area of 
110$\,$m$^2$.
The facets are made of float glass that has been slumped and polished
by the manufacturer (DOTI Technologies).
They are aluminized and anodized in a dedicated facility
near the Whipple Observatory Basecamp.
All the facets for the prototype telescope have been received;
mirror coating is in progress.
Laser measurements indicate that the optical quality of the facets
is better than originally specified:
1) the tolerance of the radius of curvature is better than
$0.4$\% (as opposed to 1.0\% specified), and
2) the average blur is less than 0.5$\,$cm (as opposed to
1.0$\,$cm specified).

\begin{figure}[t]
  \begin{center}
    \includegraphics[height=16pc]{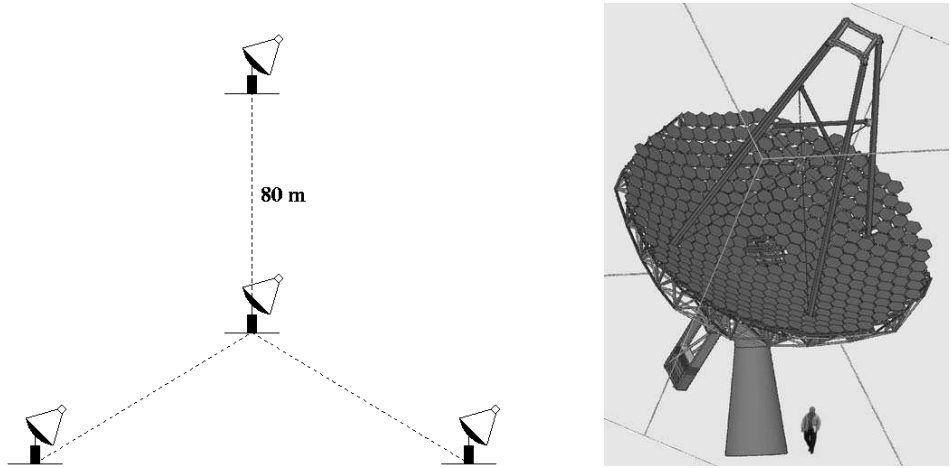}
  \end{center}
  \caption{Layout of telescopes for VERITAS-4 (left).  Three
   additonal reflectors will be added to complete the
   full array of seven telescopes.
   Design of optical support
   structure for each 12$\,$m telescope (right).}
\end{figure}

\subsection{Camera, Cables, \& High Voltage}

The cameras for VERITAS comprise 499 photomultiplier tubes (PMTs),
arranged in a close-packed hexagonal pattern with
an angular separation between PMTs of $0.15^\circ$.
Light cones increase the photon collection efficiency and protect
the PMT's from stray light pollution.
A high-speed, custom-made amplifier is used to boost the
signal before
transmission through $40\,$m
of high-bandwidth 75$\Omega$ coaxial cable.
The high voltage level for each PMT can be individually programmed, and
custom-designed electronics monitor individual PMT anode currents.

All of the key camera components were tested in November 2001 when
thirty PMT channels were installed in the focus box of the
Whipple 10m telescope.
Performance characteristics
(e.g. linearity, rise time, noise levels, etc.)
of the full electronics chain of VERITAS were verified.
The assembly of the camera for the prototype telescope is currently
in progress at the University of Chicago.  The integrated camera will be
shipped to Mt. Hopkins in spring 2003.
Figure~2 shows the camera box during assembly and an FADC trace for 
a Cherenkov pulse.

\begin{figure}[t]
  \begin{center}
    \includegraphics[height=12pc]{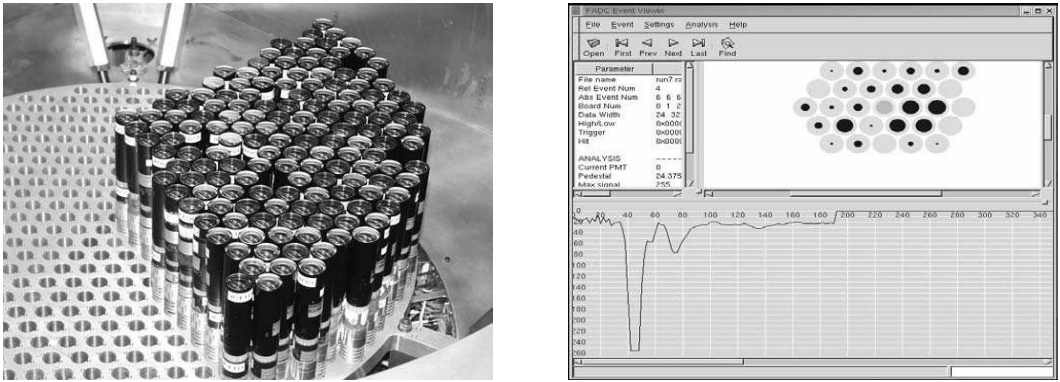}
  \end{center}
  \caption{Camera integration for prototype VERITAS telescope,
           partially completed (left).
           Display from FADC sampling of Cherenkov pulse (right).
           The second peak corresponds to a delayed copy of the pulse
           with lower gain.}
\end{figure}

\subsection{Flash ADCs (FADCs) and Data Acquisition}

The custom-built
FADCs digitize the Cherenkov pulse waveform at a rate of
500$\,$MSps to provide the maximum possible information about the
shape and time structure of the pulse.
Each PMT signal is sampled by a separate FADC with an effective dynamic
range of 11 bits and a 
memory depth of 64$\,\mu$s.
The FADCs are packaged in 9U VME boards with 10 channels/board.
The data acquisition is based around standard VME architecture,
comprising a fast VME backplane and crate Single Board
Computers (SBC's) connected to a
local event-building workstation via the fast SCI protocol.
The FADC design has been finalized, and
the boards for the prototype telescope are being built.
The VME readout for the FADC system is also near completion;
crate-to-crate transfer speeds of 100$\,$MB/s have been achieved.

\subsection{Trigger Electronics and Calibration}

VERITAS employs a three level trigger system to select
Cherenkov events at the lowest possible energy threshold.
Level 1 consists of constant fraction discriminators (CFDs)
to determine when a PMT pulse exceeds a given threshold. 
The CFDs are mounted directly on the main FADC board
to minimize noise and signal delay.
Level 2 comprises a hardware pattern trigger designed
to select compact Cherenkov events, as opposed to night sky
background overlaps. 
Level 3 uses the Level 2 telescope triggers
to determine when the array has triggered.
The designs for all three trigger levels are well advanced
and complete systems for the prototype telescope are now being
integrated with the other electronics.

The calibration system is designed to calibrate and monitor the performance
of each telescope and the combined array.
There are three major calibration components: charge injection, optical
injection, and atmospheric monitoring.
The charge injection system distributes a calibrated amount of charge to
the front-end electronics.
The initial optical system uses a dye laser 
to simultaneously flash all PMTs in the camera.

\subsection{Software}

The online software can be divided into a number of components:
1) FADC and VME data acquisition,
2) Telescope acquisition (event building),
3) Array acquisition/online analysis (Quicklook), and
4) Array Control and Database.
The software uses object-oriented (C++) programs
running on Intel-based Linux computers.
Reliance is made on widely available software packages
(e.g. CORBA, SQL, Qt, etc.) for database and
graphic user interface tasks and for
inter-process communication and control.
The majority of the online code for the prototype
telescope has been written, and now the major task is integration.
Initial software systems for offline analysis and simulation are
also in place; these will be augmented and refined in the future.

\section{Schedule \& Summary}

VERITAS is a new state-of-the-art ground-based
$\gamma$-ray observatory for VHE astronomy.
An initial phase (VERITAS-4) consists of four atmospheric
Cherenov telescopes.
A prototype telescope will begin operating in summer 2003,
and funding permitting,
first light for the VERITAS-4
array is expected in late 2005, well before the launch of 
the Gamma-ray Large Area Space Telescope (GLAST).

\section{References}

\re 1.\ Talk presented at the international symposium:
{\it The Universe Viewed in Gamma-Rays},
25-28 Sep 2002, Kashiwa, Japan.
Symposium proceedings published by Universal Academy Press, Inc,
edited by R. Enomoto.

\re 2.\ Ong, R.A. \ 2003, these proceedings.

\re 3.\ Quinn, J. {\it et al.}\ 2001,
Proc. 27th. Int. Cosmic Ray Conf. (Hamburg) 7, 2781.

\re 4.\ Weekes, T.C. {\it et al.} \ 2002,
Astroparticle Physics 17, 221.


\endofpaper
\end{document}